# PIQL: Success-Tolerant Query Processing in the Cloud

Michael Armbrust Kristal Curtis Tim Kraska
Armando Fox Michael J. Franklin David A. Patterson

AMP Lab, EECS, UC Berkeley
{marmbrus, kcurtis, kraska, fox, franklin, pattrsn}@cs.berkeley.edu

## ABSTRACT

Newly-released web applications often succumb to a “Success Disaster,” where overloaded database machines and resulting high response times destroy a previously good user experience. Unfortunately, the data independence provided by a traditional relational database system, while useful for agile development, only exacerbates the problem by hiding potentially expensive queries under simple declarative expressions. As a result, developers of these applications are increasingly abandoning relational databases in favor of imperative code written against distributed key/value stores, losing the many benefits of data independence in the process. Instead, we propose PIQL, a declarative language that also provides *scale independence* by calculating an upper bound on the number of key/value store operations that will be performed for any query. Coupled with a service level objective (SLO) compliance prediction model and PIQL’s scalable database architecture, these bounds make it easy for developers to write *success-tolerant* applications that support an arbitrarily large number of users while still providing acceptable performance. In this paper, we present the PIQL query processing system and evaluate its scale independence on hundreds of machines using two benchmarks, TPC-W and SCADr.

## 1. INTRODUCTION

Modern web development frameworks and open-source relational databases make it easier than ever for developers to rapidly turn an idea into a full-featured website. Additionally, the elasticity provided by cloud computing enables computational resources to be scaled up with only a credit card, rather than requiring huge capital investments. This agility allows developers to release new websites quickly and improve them based on continuous user feedback.

However, agility alone has proven to be no guarantee of long-term success. Imagine that due to a favorable mention on a prominent blog, a new web service quickly grows from thousands of users to millions. Driven by the number of users, the application’s database will grow by orders of magnitude in size. This rapid explosion of data often leads to growing pains in the form of increased response time or failed requests. Research has shown that even a small increase in latency has a measurable effect on user behavior [26], and websites that are unable to overcome this scaling hurdle will lose customers to their competition [25].

### 1.1 The NoSQL ‘Solution’

Modern relational database systems often exacerbate the scaling hurdle faced by successful services. As proponents of “NoSQL” solutions have widely publicized, the declarative, high-level programming interface provided by SQL database systems allows developers to inadvertently write queries that are prone to scalability problems. The issue is that data independence can hide potentially expensive operations, resulting in queries that perform well over small datasets but fail to meet performance goals as the size of the database grows. Often such performance problems are detected only after they impact site usability, and the database system provides little guidance on how to isolate and fix the problem (assuming a fix is even possible).

The scalability failure of current implementations of the traditional relational model have led many web application developers to abandon SQL-style data independence in favor of hand-coded imperative queries against distributed key/value stores. Key/value stores have been shown to provide linear scalability as machines are added and to maintain predictable per-request latencies [13]. This approach, however, creates its own problems. The use of imperative functions instead of declarative queries means that changes to the data model often require time-consuming rewrites. Perhaps more critically, developers are forced to manually parallelize requests to avoid the delays of sequential execution. Thus, the benefits of physical and logical data independence are lost.

### 1.2 Scale Independence

To address these problems, we have developed the Performance-Insightful Query Language, PIQL[1]. PIQL maintains the physical and logical data independence provided by traditional RDBMSs while introducing a new notion of data independence called *scale independence* [3]. Scale-independent queries that satisfy their performance objectives on small data sizes will continue to meet those objectives as the database size grows, even in a hyper-growth situation such as when a web service goes viral. A scale-independent system is *success-tolerant*, making it easy for developers to ensure that their implementation will be able to handle the massive onslaught of data inherent to success on the web.

Some approaches to bounding computation for interactive queries, such as GQL [2], provide a SQL-like query language but impose severe functional restrictions, such as removing joins, in order to ensure scalability. PIQL also limits queries to those that are guaranteed to scale but employs language extensions, query compilation technology, and response-time estimation to provide scale

[1]pronounced “pickle”

independence over a larger and more powerful subset of SQL. For example, as we demonstrate in Section 8, the PIQL query language is sufficiently rich to support interactive web applications, including the interactive components of the TPC-W benchmark.

## 1.3 Bounding Computation

Key to PIQL's approach to scale independence is the calculation and enforcement of bounds on the number of key/value store operations that a query will perform *regardless of the size of the underlying database*. The PIQL query compiler uses static analysis to select only query plans where it can calculate the number of key/value operations to be performed at every step in their execution. Therefore, in contrast to traditional query optimizers, the objective function of the query compiler is not to find the plan that is fastest on average. Rather, the goal is to avoid performance degradation as the database grows. Thus, the compiler will choose a potentially slower bounded plan over an unbounded plan that happens to be faster given the current database statistics. If the PIQL compiler cannot create a bounded plan for a query, it warns the developer and suggests possible ways to bound the computation.

This static analysis can be performed for some queries using existing annotations, such as the `LIMIT` clause [9] or foreign key constraints. However, in many cases, it is insufficient to simply limit the result size as intermediate steps also contribute to execution time. Therefore, PIQL extends SQL to allow developers to provide extra bounding information to the compiler. First, PIQL provides a `PAGINATE` clause, allowing the results of unbounded queries to be efficiently traversed, one scale-independent interaction at a time. Second, PIQL enables bounding intermediate results through relationship cardinality constraints in the database schema.

## 1.4 Meeting SLOs

The bounded number of storage system operations is the dominant driver of cost in PIQL query execution. However, simply having an upper bound on the number of key/value store operations is not enough to ensure customer satisfaction because, for interactive applications, performance objectives are typically based on response time rather than operation count. In this paper, we focus on applications whose performance requirements are expressed in terms of Service Level Objectives (SLOs) framed as a target response time for a fraction of the queries observed during a given time interval; e.g., "99% of queries during each ten-minute interval should complete in under 500 ms."

PIQL provides an SLO compliance prediction model that uses the query plan and the operation bounds to calculate the likelihood of a (scale-independent) PIQL query meeting its SLO. In Section 8, we show that for our benchmark queries, even a simple model can accurately predict SLO compliance.

## 1.5 Summary

In this paper, we describe the PIQL language and system components that implement and extend the original vision proposed in an earlier position paper [4]. We demonstrate the expressiveness of the PIQL language, as well as the scale independence of our implementation, using two benchmarks: TPC-W, an online store, and SCADr, a simplified microblogging service. We show linear increases in request throughput and validate our SLO compliance model on clusters of up to 150 machines. In summary, this paper contains the following contributions:

- We describe the notion of scale independence for supporting web applications in a "success-tolerant" manner and outline an approach for achieving scale independence without sacrificing data independence.
- We present PIQL, a minimal extension to SQL that allows developers to express relationship cardinality and result size requirements.
- We describe the PIQL query compiler, which bounds the number of key/value store operations performed for a given query.
- We present a performance model that helps developers determine acceptable relationship cardinalities and reason about SLO compliance.
- We demonstrate the expressiveness of the PIQL language, the accuracy of our SLO compliance prediction, and the scale independence of our implementation using two benchmarks.

The remainder of this paper is organized as follows: Section 2 describes different classes of queries and how their performance relates to the total size of the database. Section 3 presents the overall architecture of the PIQL database engine, followed by a description of the DDL and DML extensions in Sections 4, the scale-independent optimization techniques in Section 5, the prediction framework in Section 6, and the execution engine in Section 7. In Section 8, we summarize the results of our experiments using the TPC-W and SCADr benchmarks. Section 9 discusses related work. Section 10 presents conclusions and future research challenges.

# 2. QUERY SCALING CLASSES

Before going into the details of the PIQL system, it is useful to step back and consider the sources of scale *dependence* in interactive web applications. As shown in Figure 1, we can divide queries into classes based on their performance scalability as the database size increases. We briefly describe each of these classes below.

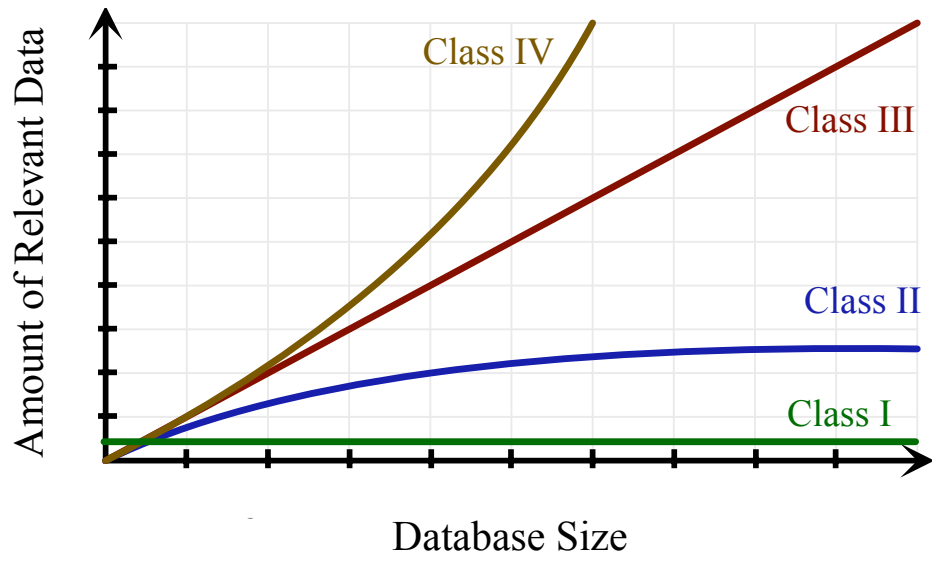

**Figure 1: A comparison of the scalability of various queries as database size increases.**

**Class I (Constant):** In the simplest case, the amount of data required to process a query is constant. For example, in a web shop, data needed to display a particular product or to show the profile of a particular user based on a unique ID is naturally limited regardless of how many products or users there are in the database. The optimizer knows about this bound due to the fact that such a query would have an equality predicate against the primary key of the relations. Other types of queries that fall into this class include queries with a fixed `LIMIT` that do not perform any joins, or that only perform joins against a unique primary key.

**Class II (Bounded):** A second class of query involves data that will grow as the site becomes more successful but that is naturally bounded. For example, in social networks, it is known that while people will gradually add more friends over time, the average person has around 150 "real" friends [18]. Setting a maximum friend limit of 5000 friends, as is done by Facebook, satisfies most customers [7]. PIQL allows the developer to express these limits explicitly in the schema, through an extension to the DDL (see Section 4.2).

**Class III (Sub-linear or Linear):** These queries require touching an amount of data that grows sub-linearly or linearly as the site becomes more successful. A carelessly-written query listing all currently logged-in users or a count over all customers falls into this category.

**Class IV (Super-linear):** These queries require computations over intermediate results that grow super-linearly with the number of users. For example, clustering algorithms that require computation of a self Cartesian product would fall into this class.

By definition, a success-tolerant web application can support only queries from Classes I and II. PIQL identifies such queries and, in the case of Class II queries, provides hints of acceptable cardinality constraints for meeting specific SLOs. For Class III and IV queries, PIQL can identify portions of queries that are unbounded and can suggest workarounds, such as the introduction of cardinality constraints or the use of pagination.

# 3. ARCHITECTURE OVERVIEW

To achieve scale independence, PIQL is designed to leverage a distributed key/value store through a library-centric database architecture. PIQL uses the key/value store as a record manager and provides all higher-level functionality (such as a declarative query language, relational execution engine, and secondary indexes) via a database library. This approach is similar to the architecture employed by Google's Megastore, as well as by Brantner et al. [6, 8].

In this architecture (as shown in Figure 2), each application server includes a PIQL database engine library that directly communicates with the key/value store. In accordance with best practices, the application servers and thus the database library are designed to avoid preserving state between requests. This separation of the database into a stateless component (the database library) and a stateful component (the key/value store) decreases the complexity of our system significantly. Query processing is performed at the client, thereby minimizing the functionality of the stateful component. As a result, scaling and load distribution can be performed using standard key/value store techniques [13].

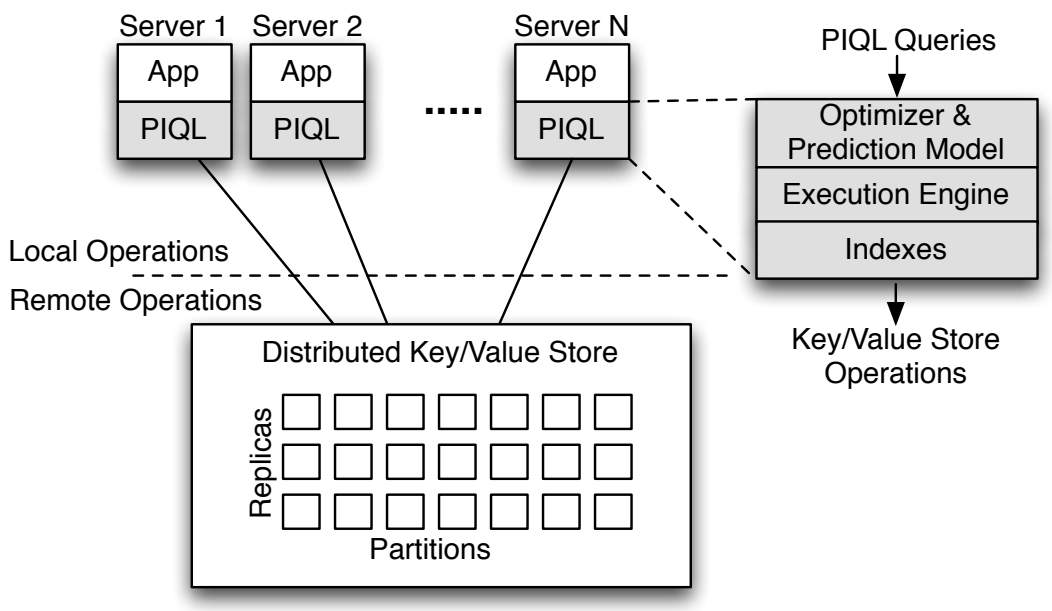

**Figure 2: The PIQL database engine is implemented as a library that runs in the application tier and communicates with the underlying key/value store.**

PIQL maintains predictable response time by building on the predictable performance provided by modern key/value stores. For example, Amazon's Dynamo [13] demonstrated consistent performance for get/put operations even in the 99.9th percentile on a large commodity cluster during their peak shopping season (December 2006). Of course, a web application may experience load spikes in addition to normal fluctuation due to diurnal/seasonal usage patterns. However, recent research [28] addresses this situation using an approach based on control theory in addition to well-known best practices such as replication and over-provisioning. Their experiments make use of the elasticity available in a cloud environment to scale a key/value store up or down in response to load changes while maintaining SLO compliance.

PIQL also requires that the key/value store supports range requests in order to provide data locality during index scans. Many key/value stores already fulfill this requirement by using range-preserving index structures [10, 12].

Finally, PIQL relies on the key/value store to provide consistency. It is important that the mechanism used to provide consistency is nonblocking, as the variable performance that results from blocking operations such as waiting for a lock would violate the requirement of predictable performance for key/value store operations. Our prototype currently meets the non-blocking requirement by implementing eventual consistency. In Section 7.2, we provide more detail about the consistency semantics of our system. Alternatively, if a given application requires stronger consistency, it can be achieved using other non-blocking techniques such as snapshot isolation. A full discussion on how higher-level consistency guarantees can be implemented on top of key/value stores can be found in [8, 22].

# 4. PIQL LANGUAGE EXTENSIONS

As stated in the introduction, PIQL extends standard SQL with constructs that allow developers to bound not only the number of results returned for each user interaction with the database, but also to limit the cardinality of intermediate results. In this section, we describe the DDL and DML extensions that enable the PIQL query compiler to bound the number of operations required even for complex queries involving joins or unbounded amounts of data.

## 4.1 Bounding Data Returned

In many applications, there are cases where the developer needs to run queries that would return potentially unbounded amounts of data, such as a query that lists all posts made by a user in chronological order. While this query itself cannot be made scale-independent, PIQL has a language feature that allows the developer to bound the number of key/value store operations required for each user interaction with the database by displaying subsets of the full result one page at a time.

PIQL queries can contain a `PAGINATE` clause which specifies how many items should be returned for each user interaction with the database. Paginated queries are implemented as client-side cursors and can be invoked repeatedly, returning the next page of results each time. PIQL also supports the more traditional `LIMIT` clause for cases where only the top-K results of the query are required.

Additionally, to simplify the request routing and preserve the stateless nature of the application servers, the client-side cursor can be serialized and shipped to a user along with the results of the query. When the next page is desired, the serialized state is sent back to any application server where it is deserialized, and execution can be resumed. The size of a serialized client-side cursor is generally small as we only need to remember the last key returned by any uncompleted index scans in the query.

Note that the traditional methods of implementing pagination either require onerous server-side state management or are not scale-independent. Specifically, using server-side cursors requires the maintenance and garbage collection of the cursor state, even in the face of hundreds or thousands of users coming and going. The other common implementation of pagination uses both `OFFSET` and `LIMIT` clauses. Unfortunately, executing a query with an offset requires work proportional to the size of the offset, which is in conflict with our goal of scale independence [2].

## 4.2 Bounding Intermediate Results

Standard SQL referential integrity constraints in the schema definition already allow the compiler to infer cardinality in one direction, from a foreign key to a single corresponding tuple. PIQL extends these constraints by allowing the expression of relationship cardinalities in the other direction as well. These developer-specified relationship cardinalities provide extra information for the optimizer and execution system about natural limits to the various relationships; these limits are often due to real-world constraints (see Section 2). The form of this specification is a maximum number of tuples that may contain a distinct value or set of values. For example, in our sample microblogging application, SCADr, a limit is placed on the number of users to whom a single user may subscribe. This limit is expressed in the following schema:

```
CREATE TABLE Users (
  userId INT,
  firstName VARCHAR(255)
  ...
)
CREATE TABLE Subscriptions (
  ownerUserId INT,
  targetUserId INT,
  ...
  CARDINALITY LIMIT 100 (ownerUserId)
)
```

By specifying that there is a limit of 100 on the cardinality of any specific value of `ownerUserId` in the Subscriptions table, the developer informs the optimizer that no single user is allowed to have more than 100 subscriptions. Section 5 discusses in more detail how this limit is used during optimization.

Choosing an appropriate limit is crucial. As mentioned earlier, Facebook decided to use a very loose limit, 5000, for the number of friends (recall that the natural limit is closer to 150). This caused some power users with more than 3000 friends to complain about the response time [27], as the system significantly slows down and some features completely break. In contrast, a Facebook competitor, Path, limits the number of friends to 50, which is even smaller than the natural limit. With PIQL's prediction framework, we offer a tool to determine acceptable limits so that all queries meet the SLO requirements, independent of the scale of the system (see Section 6).

## 5. SCALE-INDEPENDENT OPTIMIZATION

Given a query expressed in PIQL, the optimizer must select a scale-independent ordering of physical operators for its execution. PIQL's optimizer operates in two phases as sketched in Algorithm 1 and Algorithm 2. As a concrete example, consider the thoughtstream query from SCADr, which allows a user to retrieve the most recent "thoughts" (in Twitter terminology, "tweets") of all the users to whom they are currently subscribed. Figure 3 shows the phases of optimization performed on the thoughtstream query.

In the following, we describe the algorithm in more detail. We concentrate on the scale-independent aspects of the query optimizer. Other optimizations, such as selecting join orderings, are performed using traditional techniques (e.g., [23]) and are therefore not discussed.

### 5.1 Phase I: Stop Operator Insertion

Given a logical plan from the query parser, Phase I starts by finding an appropriate linear join ordering (Line 1 in Algorithm 1). Next, the optimizer pushes predicates down in the plan using standard techniques (Line 2).

**Algorithm 1** StopOperatorPrepare - Phase I

**Require:** $logicalPlan \leftarrow$ Logical PIQL Plan
**Require:** $cardinalityConstraints \leftarrow$ Developer-Specified Cardinality Constraints
1: $orderedPlan \leftarrow findLinearJoinOrdering(logicalPlan)$
2: $preparedPlan \leftarrow predicatePushDown(orderedPlan)$
3: **for all** relation $r$ in $preparedPlan$ **do**
4:   **for all** combinations $c$ of AttributeEquality predicates against $r$ **do**
5:     **if** $attributesOf(c)$ contains all fields in $primaryKey(r)$ **then**
6:       Insert *data-stop* of cardinality 1 above $c$
7:     **else if** $attributesOf(c)$ contains all fields of a cardinality constraint **then**
8:       Insert *data-stop* of specified cardinality above $c$
9:     **end if**
10:   **end for**
11: **end for**
12: $finalLogicalPlan \leftarrow stopPushDown(preparedPlan)$
13: **return** finalLogicalPlan

Due to a `LIMIT` or `PAGINATE` clause in the actual query, the logical plan might already contain a standard *stop* operator to restrict the number of tuples returned [9]. Additionally, the optimizer will introduce new *data-stop* operators to the logical plan where schema-based cardinality information exists (Lines 3 to 11). The data-stop operator is a new operator that acts as an annotation, telling Phase II of the optimizer that a given section of the plan will produce no more than the specified number of tuples due to a schema cardinality constraint. Any time equality predicates reference the entire primary key of the relation, a data-stop operator is inserted into the plan with a cardinality of one (Lines 5-6). Otherwise, if equality predicates reference all of the fields in a `CARDINALITY LIMIT`, a data-stop operator is inserted into the logical plan with the given cardinality (Lines 7-8).

Afterwards, data-stop operators from the insertion phase as well as stop operators from a `LIMIT` or `PAGINATION` clause are pushed down as deep as possible into the plan (Line 12). To ensure the query can be executed without restart, the stop operators start at the top of the plan and are pushed down conservatively according to rules regarding non-reductive predicates [9]. Specifically, a stop operator cannot be pushed past a predicate that might reduce the number of tuples, as this could lead to an incorrect plan that produces fewer tuples than requested.

Recall that data-stop operators are hints which are inserted based on the number of tuples that can possibly be stored in the database as enforced by the DDL constraints, instead of the number of results desired for the query. Therefore, a data-stop operator can be pushed past all predicates other than those that caused its insertion. This push-down is possible because even if a predicate reduces the number of tuples produced by the query, there cannot be any more tuples in the database due to the cardinality constraint.

This flexibility allows the data-stop operator to be pushed further down in the plan, making scale-independent static analysis of more queries possible. For example, in the optimization of the thoughtstream query, the data-stop operator is pushed past the predicate that ensures a given subscription was approved. This would not have been possible with a standard stop operator. Since the optimizer is able to bound this section of the plan, its heuristic then chooses a local selection against the primary index instead of creating a new index that includes the approval field. This is cheaper both because it avoids maintaining an unnecessary index and because lookups over an index require an extra round trip to the key/value store to retrieve the full tuple.

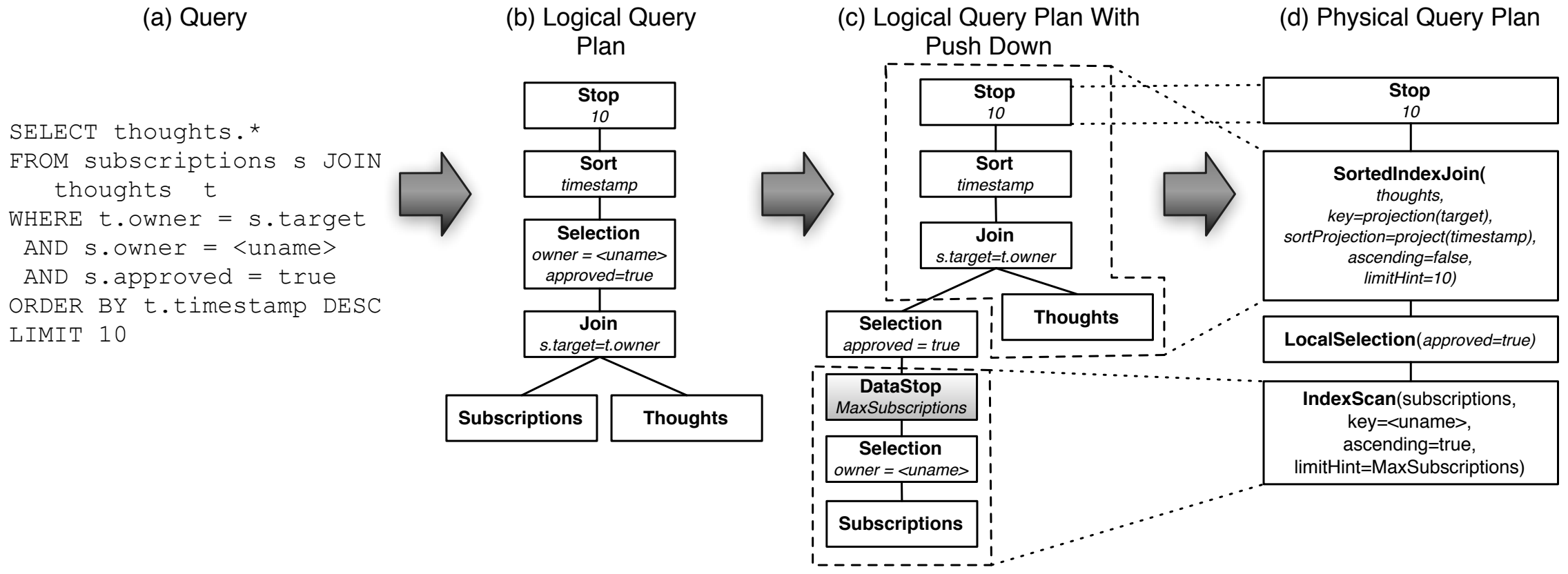

**Figure 3: The stages of optimization for the thoughtstream query in SCADr.**

## 5.2 Phase II: Physical Operator Selection

After the predicate and stop operator push-down, the optimizer transforms the logical plan recursively into a physical plan (Algorithm 2). The physical operators of the PIQL execution engine are broken into two groups: those that operate locally on the client executing the query, and those that issue requests to the key/value store.

### 5.2.1 Remote Operator Matching

In order to ensure scale independence, the optimizer requires each remote operator in the plan to have an explicit bound. This means that whenever a plan section contains a group of logical operators that will be mapped to a remote operator, it must have either a stop operator or a foreign key uniqueness constraint. This requirement ensures that there will be a bound not only on the final result set, but also on any intermediate results that must be shipped across the network from the storage tier to the query processing library.

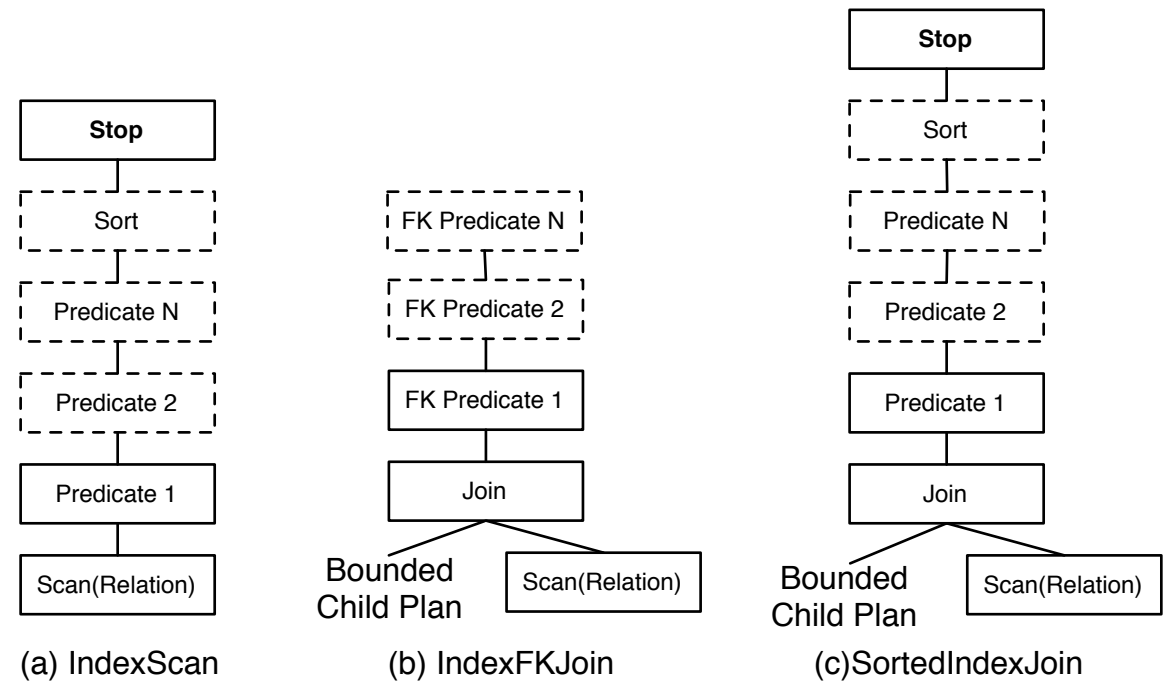

**Figure 4: Every remote operator is equivalent to a pattern of logical operators. Optional logical operators are denoted by dotted line boxes.**

Our current system supports three remote operators: *IndexScan*, *IndexFKJoin*, and *SortedIndexJoin*. Figure 4 shows the groups of two or more logical operators that are equivalent to a given remote operator. In the following, we describe the rules for mapping logical operators to the three remote operators.

**Index Scan** maps to a set of predicates evaluated against a relation where the predicates describe some contiguous section of the index. Figure 4 (a) shows all the logical operators a single index scan can cover. In practice, this means that there can be any number of equality predicates, while predicates involving inequality may touch at most one attribute. Additionally, an index scan can be used to satisfy a logical sort operator using the special ordering of the index. However, if there is an attribute involved in an inequality, it must be the first field of any sort order to be satisfied by the index scan; otherwise, this collection of predicates and sort constraints would by definition describe a potentially non-contiguous section of the index. Attempting to return potentially non-contiguous tuples from an index scan would make it impossible to bound the amount of work required to produce a given number of matching tuples.

**Index Foreign Key Join** maps to a join where the predicates constitute equality for the primary key of another table. Due to the uniqueness constraint of the primary key, the optimizer knows that the resulting number of tuples produced by the join will be less than or equal to the number of tuples produced by the child plan and will thus be bounded in size. The plan coverage is shown in Figure 4 (b).

**Sorted Index Join** maps to a join where there is an optional sort before the next available *limit hint* shown in Figure 4 (c). By using a composite index, the tuples can be pre-sorted for every join key and thereby leverage the knowledge of the limit hint to bound the number of data items per join key. For example, the thoughtstream query of Figure 3 (c) would normally require all thoughts per subscription. However, by pre-sorting the thoughts per subscription, the operator is able to receive only the latest limit hint thoughts (here 10) per subscription, which enables the overall bound.

### 5.2.2 Local Operator Matching

Local operators (i.e., operators that run in the application tier) include sort, select, group by, and various aggregates. In contrast to remote operators, local operators work entirely on local data, which is shipped to the client. Consequently, as the remote operators ensure that all data is bounded in size, all local operators are bounded as well. The query language does not allow recursion; therefore, it is impossible for the local result size to increase infinitely.

### 5.2.3 Physical Plan Generation Algorithm

The general algorithm to transform the logical plan to a physical one is shown in Algorithm 2. Starting from the top of the logical plan, the compiler tries to map as many logical operators as possible to a bounded physical remote operator according to the rules of Section 5.2.1 (Line 1). If the compiler finds a remote operator, it recursively calls the generator function with the remaining logical plan and attaches the resulting optimized child plan (Line 2-6). If no remote operator can be found, the compiler tries to find a local operator (Line 7) and if successful, it continues recursively (Line 8). If all logical operators are successfully matched either to a remote or local operator, a bounded query plan is found and returned (Line 10). However, if at any stage it is impossible to find either a remote or local operator, the plan is assumed to be unbounded and is therefore not scale-independent (Line 12).

Figure 3 (c)-(d) shows an example transformation from a logical plan to a physical plan. The algorithm first selects an IndexScan to retrieve the subscriptions for a given owner. Note that this selection was made possible by the data-stop operator inserted as a result of the cardinality constraint on the number of subscriptions allowed for a given owner. Afterwards, it chooses a LocalSelection on the approved status, followed by a SortedIndexJoin and stop operator.

**Algorithm 2** PlanGenerate - Phase II

```
Require: logicalPlan ← Logical PIQL Plan
 1: if (remoteType, child) ← match remote operator then
 2:     if logicalPlan has standardStopOperator then
 3:         return stop(remoteType(PlanGenerate(child)))
 4:     else
 5:         return remoteType(PlanGenerate(child))
 6:     end if
 7: else if (localType, child) ← match local operator then
 8:     return localType(PlanGenerate(child))
 9: else if logicalPlan = Ø then
10:     return Ø
11: else
12:     ERROR(Not scale-independent)
13: end if
```

## 5.3 Index Selection

Since table scans are not scale-independent (they might fall into Class III from Section 2), the PIQL optimizer produces a list of all necessary indexes during query optimization. These indexes can be automatically created by the system. For example, consider the following query from the TPC-W benchmark:

```
SELECT I_TITLE, I_ID, A_FNAME, A_LNAME
FROM ITEM, AUTHOR
WHERE I_A_ID = A_ID
  AND I_TITLE LIKE [1: titleWord]
ORDER BY I_TITLE
LIMIT 50
```

The PIQL optimizer will select an IndexScan over an index consisting of the fields (token(I_TITLE), I_TITLE, I_ID) with a limit hint of 50. The first field allows the IndexScan to find all of the titles that contain the given token. The second field ensures that the items returned by taking the top 50 records from this index will be sorted by the full title of the item. Finally, the I_ID allows the execution engine to dereference the index and retrieve the actual item. Above this IndexScan, the optimizer would place a join with the author relation on the primary key A_ID.

# 6. SLO COMPLIANCE PREDICTION

In the previous sections, we covered how the PIQL compiler converts a PIQL query into a scale-independent query plan. If PIQL is unable to find such a plan, the query is reported to the developer as a possible performance risk at scale through the Performance Insight Assistant (see Section 6.4). Even if the compiler is able to find a bounded plan, however, it still does not guarantee that the plan is success-tolerant (i.e., that it can be executed in the targeted latency time frame). The number of tuples to process, although bounded, might still be too high to meet response-time objectives.

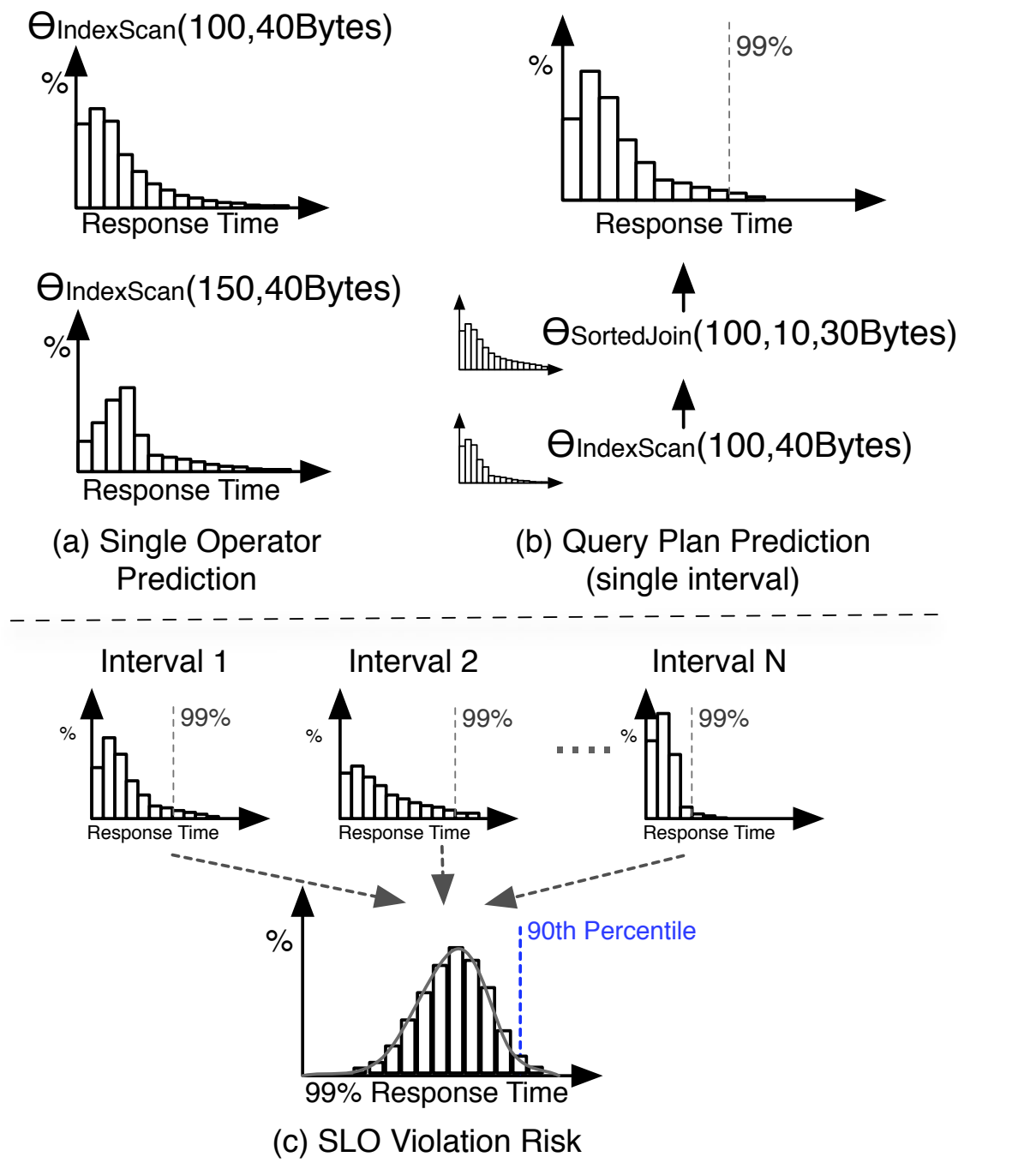

**Figure 5: Modeling process for PIQL queries. First, we create models for each PIQL operator (a). Then, for each query, we combine the operator models according to the query plan to create a PDF for the whole distribution (b). Finally, we repeat the process of (b) for many timeframe histograms to better reflect the SLO response-time risk (c).**

In this section, we describe the response time prediction model used by PIQL. This model calculates the risk that query operations will not complete in the targeted time frame. Using this prediction model, the developer will be informed at compile time whether a bounded query is likely to meet its SLO. In the remainder of this section, we first describe how we model a single query plan operator and then how we compose operators together to evaluate the response time of the whole query as well as the risk of violating the SLO. Finally, we describe how the model is used as part of the Performance Insight Assistant.

## 6.1 Single Operator Model

As described in the previous section, a physical query plan is composed of one or more operators. To reflect the volatility in the response time, we model each operator as a random variable $\Theta$. We assume that the response-time distribution of an operator only depends on the number of tuples and the size per tuple it has to process. This simplification is reasonable, as our architecture is designed to avoid contention by automatically load-balancing and re-provisioning the key/value store as well as the application tier (see Section 3). We further simplify the model by only considering the three remote operators, which interact with the key/value store. This simplification is sufficient, as we only target interactive queries with SLO goals in the range of milliseconds up to a few seconds, which makes the latency to the key/value store the dominating factor. However, network bandwidth and round-trip times are improving. Thus, in future versions we plan to extend the model to include the local operators as well.

Accordingly, the IndexScan operator can be modeled as $\Theta(\alpha, \beta)$, where $\alpha$ represents the number of expected tuples (i.e., the limit hint) and $\beta$ the size per tuple. In contrast, the two join operators are described as $\Theta(\alpha_c, \alpha_j, \beta_j)$, where $\alpha_c$ represents the maximum number of tuples from the child operator (right relation), $\alpha_j$ the bound (e.g., schema cardinality) between the left and right relations and $\beta_j$ the maximum size of a tuple from the left relation. We do not need to consider the size of the child tuples, as those tuples are already local during query processing.

In our current model, we obtain an empirical distribution for each of the operator random variables and store it as a histogram. That is, as part of the model training, we sample the response time behavior for every operator by repeatedly executing the operator with varying cardinality and tuple sizes. This training is typically done once by setting up a production system in the cloud for a short period of time (see Section 8), where we measure all operators in parallel. We use this sampling to collect a set of histograms for our three remote operators with different $\alpha$ and $\beta$ values. Since these statistics are not application-specific, they could even be pre-calculated for the most prominent public clouds (e.g., Amazon, Google, Microsoft). Figure 5(a) shows two possible distributions for the IndexScan operator with an expected cardinality limit $\alpha$ of 100 and 150 and a tuple size $\beta$ of 40 Bytes. The models can be updated periodically as conditions in the datacenter change (e.g., as hardware is upgraded).

Given a query plan, we can obtain the maximum cardinality $\alpha$ and the maximum tuple size $\beta$ for each operator from the optimizer annotations and the schema, respectively. Thus, choosing a distribution from the histogram collection becomes as simple as looking up the correct $\alpha$ and $\beta$, which we set to be the maximum cardinality to avoid underestimating the response time. If the correct values are not in the table, we can choose the $(\alpha, \beta)$ setting that is closest to the desired value while still being larger. For example, for the IndexScan with a username on the *Subscriptions* table described in Section 4.2, we know from the schema annotation that the cardinality is 150 and the tuple size is 40 Bytes. Hence, we would choose $\Theta_{IndexScan}(150, 40\text{Bytes})$ from the two histograms in Figure 5(a). In the future, we also plan to explore interpolating among the stored models to produce the desired model; this technique is suitable since our system exhibits a linear relationship between the cardinality and the response time.

Since our system is designed for interactive applications, the response time goals for all queries are typically less than one second. For these purposes, we believe that reporting values with millisecond resolution is sufficient, so each histogram can be well-represented with on the order of a thousand bins. Therefore, while it is true that our approach requires a separate histogram per $(\alpha, \beta)$ pair, this burden is not onerous; due to the limited resolution of interest, each histogram can be stored in a kilobyte or two.

## 6.2 Query Plan Model

To predict the overall query response time, we combine the operator models according to the physical query plan generated by the PIQL optimizer. Here, we can make use of another property of our architecture. Our execution engine is implemented as an iterator model and thus allows executing several operators in a pipelined fashion; however, since we restrict our attention to short-running queries, the latency can be modeled with sufficient accuracy assuming blocking operators. In the worst case, the model fails to capture the overlap among the operators, and our prediction is overly conservative. However, recall that our goal is not to predict response time but rather SLO compliance; thus, as long as the prediction is below the SLO, it still correctly predicts SLO compliance (see Section 8.6 for quantitative analysis).

Accordingly, we simplify our model by assuming independence among the operators. For query plans (or plan sections) that are serial, we represent the overall latency with a random variable whose latency is the sum of the operator latencies, each of which is also represented by a random variable. For parallel plan sections, e.g. the two child plans of a union operator, we determine the latency of each branch and then take the maximum. Since we view the latency of each operator as a random variable, summing the latency of two operators is equivalent to convolving their densities. Thus, to predict the latency distribution of a query, we convolve the densities of its operators, and the resulting distribution is that of the query, shown in Figure 5(b). Recall that we ignore the local operators, as requests to the key/value store dominate the latency. Accordingly, modeling the timeline query of SCADr shown in Figure 3 requires convolving two operators:

$$Q_{ThoughtStream} = \Theta_{IndexScan}(SubscrCard, SubscrSize) * \Theta_{SortedJoin}(SubscrCard, ThoughtsCard, ThoughtSize)$$

## 6.3 Modeling the Volatility of the Cloud

Our goal is to determine whether a query will meet its SLO regardless of the underlying database size; to do so, we inspect its predicted latency distribution. We are chiefly concerned with detecting violations of SLOs that are defined in terms of high quantiles of the query latency distribution; thus, given an SLO like "99% of queries during each ten-minute interval should complete in under 500 ms," if the 99th-percentile latency of our predicted distribution is less than 500 ms, we predict that the query will meet the SLO. Note that the length of the SLO interval impacts its stringency; longer intervals make the SLO easier to meet, as any brief periods of poor performance are counterbalanced by mostly good performance. In what follows, we assume the SLOs are defined over non-overlapping time intervals.

The 99th-percentile latency can vary from one interval to the next, which poses a new challenge for the model. As mentioned in Section 3, we assume that the key/value store's performance is relatively stable. Natural fluctuations in performance are particularly common in public clouds, where the machines and network are shared among many clients. Heavy workloads of some clients (e.g., Netflix's video encoding on Amazon) might cause short periods of poor performance, which could result in violations to an SLO even though it is routinely met under normal operation. Therefore, rather than providing a point estimate for the 99th-percentile latency of a given query, we estimate its distribution, which captures how it varies from one interval to the next. In order to estimate a distribution, we take the data collected from benchmarking the operators and bin the data according to the interval of interest; e.g., if the SLO is provided over a ten-minute interval, we create a separate histogram for each ten-minute period. This process allows us to obtain a prediction of the query's 99th-percentile latency for each interval during which the benchmark was observed. Combining these predictions, as in Figure 5(c), we obtain a prediction of the distribution of the 99th-percentile latency. This distribution is a useful tool to a developer as it provides information about the risk of violating a query's SLO over time. For example, if the target response time equals the 90th percentile of the distribution, it means that for 10% of the intervals considered, the SLO goal may be violated.

## 6.4 Performance Insight Assistant

In order to make it easier for a developer to work within the constraints enforced by the PIQL optimizer, the system provides helpful feedback for fixing "unsafe" queries and for appropriately sizing cardinality limitations. Regarding the first case, any time a query is rejected by the optimizer, the developer is provided with a diagram of the logical query plan where the problematic segment is highlighted. The system provides the developer with possible attributes where the addition of a `CARDINALITY LIMIT` would allow optimization to proceed.

| Number of Subscriptions per User \ Records per Page | 10 | 15 | 20 | 25 | 30 | 35 | 40 | 45 | 50 |
|---|---|---|---|---|---|---|---|---|---|
| 100 | 301 | 272 | 303 | 310 | 324 | 327 | 350 | 355 | 349 |
| 150 | 348 | 362 | 390 | 401 | 417 | 421 | 435 | 467 | 484 |
| 200 | 411 | 452 | 460 | 479 | 512 | 523 | 547 | 573 | 604 |
| 250 | 496 | 511 | 563 | 559 | 600 | 623 | 659 | 686 | 709 |
| 300 | 556 | 602 | 625 | 645 | 680 | 722 | 773 | 797 | 829 |
| 350 | 636 | 687 | 700 | 744 | 772 | 820 | 848 | 887 | 928 |
| 400 | 679 | 726 | 776 | 835 | 853 | 882 | 936 | 953 | 976 |
| 450 | 766 | 815 | 861 | 902 | 914 | 948 | 971 | 980 | 987 |
| 500 | 832 | 887 | 934 | 943 | 973 | 978 | 987 | 985 | 987 |

**Figure 6: Predicted heatmap for 99th percentile latency (ms) for the thoughtstream query. On average, the predicted values are 13 ms higher than the actual values.**

For example, recall the thoughtstream query presented in Figure 3. If the developer had not specified a limit on the number of subscriptions that a user could own, optimization would have failed. The assistant would have then pointed out that the problem was with the number of tuples produced by the subscription relation. The developer should then set a limit on the number of subscriptions per owner.

The performance insight assistant also provides guidance on how to set cardinality limits that are compatible with SLO compliance. Given a query written in PIQL, the system will provide the developer with a chart that shows how the 99th-percentile response time will vary for different cardinality limits; this chart is obtained by predicting the latency distribution for each setting of the cardinality using our per-operator benchmarks. If the developer specifies an SLO, the system can suggest values that maximize functionality while still meeting performance requirements.

Figure 6 shows this analysis performed for SCADr's thoughtstream query. The thoughtstream query has two parameters: the number of subscriptions a user has and the total number of thoughts to return. Since this query has two parameters, choosing the cardinality limits for this query is more complicated. Thus, we provide a heatmap so that the developer can see what the latency would be for each combination of the parameters. The developer can choose any of the cardinality pairs that would satisfy the query's SLO.

Averaging over all of the considered (number of subscriptions, number per page) pairs, the predicted values are 13 ms higher than the actual measured values. Section 8.6 contains a complete query-by-query evaluation of our predication accuracy.

# 7. EXECUTION ENGINE

Much of PIQL's database engine is implemented using standard techniques. There are, however, several notable differences that arise as a result of the performance characteristics and limitations of the underlying key/value store.

## 7.1 Physical Operators

The physical operators are implemented using an iterator model [17]. The iterator interface provides open(), next(), and close() methods. The implementation of the local operators is very similar to their counterparts in a traditional system.

In contrast, since remote operators retrieve data by issuing relatively high-latency requests against the key/value store, the trade-off between lazy evaluation and prefetching is even more severe than in a traditional system. Additionally, a key/value store, unlike a record manager running on a single machine, can support many requests made in parallel without interference. This enables two optimizations: First, the execution engine can leverage the limit hint information from the compiler to prefetch all required data in a single request. Second, each of the operators is able to issue all of the requests to the key/value store in parallel. Our experimental results, presented in Section 8, show that batching and parallelism greatly reduce the total response time for a given query. The current implementation calculates all group-bys and aggregates in the client tier and thus requires that they are only performed on datasets that are known to be bounded.

## 7.2 Index Management and Consistency

Since the underlying key/value store used in our prototype supports only eventual consistency, our prototype must build on the primitives provided, for instance test-and-set, to deliver the expected consistency semantics of the benchmarks. For example, maintaining secondary indexes requires a form of atomicity, as a crash might cause indexes to never be updated and thus not even provide eventual consistency. In our prototype, we avoid this potential inconsistency by first inserting all new keys to all secondary indexes, then updating the record, and finally deleting all stale secondary index entries. Doing so might at most cause the index to contain some dangling pointers, which can be garbage-collected.

Additionally, our prototype ensures the cardinality constraints on relationships using the following protocol: after inserting an item, the system checks the cardinality constraint using a count range request, or a range request if count is not supported. If the total count returned is less than the constraint, the insert is considered successful. If not, the inserted record is deleted. Note that this protocol might temporarily violate cardinality constraints for concurrent insertions. Finally, our prototype supports uniqueness constraints and conditional updates through the test-and-set operator.

## 7.3 Wildcard Lookups

Evaluating arbitrary regular expressions over an ever increasing amount of data is not scale-independent. Our prototype instead supports searching for a given token through the use of an inverted full-text index. More complicated substring matching could also be supported without hindering scale-independence if the key/value store implemented an index that allows substring lookups in constant time such as a distributed suffix tree [11].

# 8. EXPERIMENTS AND RESULTS

We evaluated our prototype using two benchmarks: one based on the user-facing queries of TPC-W and another called SCADr. TPC-W models a typical web shop, like Amazon.com. SCADr simulates a website similar to the microblogging platform Twitter. We implemented both benchmarks in order to evaluate the expressively of PIQL and executed them on up to 150 Amazon EC2 nodes running the SCADS key/value store [3] to verify the scale independence of our system and the accuracy of our SLO compliance prediction. This section describes the two benchmarks and their setup in more detail and then reports on our results.

| | Query | Modifications | Additional Indexes | Actual 99th (ms) | Predicted 99th (ms) |
|---|---|---|---|---|---|
| TPC-W Benchmark | Home WI | - | - | 94 | 95 |
| | New Products WI | Tokenized search | Items(Token(I_SUBJECT), I_PUB_DATE) | 302 | 395 |
| | Product Detail WI | - | - | 118 | 125 |
| | Search By Author WI | Tokenized search | Authors(Token(A_FNAME, A_LNAME)), Items(I_A_ID, I_TITLE) | 138 | 136 |
| | Search By Title WI | Tokenized search | Items(Token(I_TITLE), I_TITLE, I_A_ID) | 122 | 145 |
| | Order Display WI Get Customer | - | - | 97 | 95 |
| | Order Display WI Get Last Order | - | Orders(O_C_UNAME, O_DATE_TIME) | 176 | 207 |
| | Order Display WI Get OrderLines | - | - | 126 | 138 |
| | Buy Request WI | - | - | 130 | 148 |
| SCADr | Users Followed | - | - | 113 | 141 |
| | Recent Thoughts | - | - | 88 | 89 |
| | Thoughtstream | Cardinality constraint on #subscriptions | - | 140 | 153 |
| | Find User | - | - | 84 | 82 |

**Table 1: The query modifications and indexes required for scale-independent execution of SCADr and TPC-W, as well as predicted and actual 99th-percentile response times.**

## 8.1 Benchmarks

### *8.1.1 TPC-W Customer Queries*

The TPC-W benchmark is a throughput benchmark for database web applications. It models an online bookstore with a mix of fourteen different kinds of requests such as searching for products, displaying products, and placing an order. Every request consists of one or more queries to render the corresponding web page. Furthermore, the TPC-W benchmark specifies three kinds of workload mixes: (a) browsing, (b) shopping, and (c) ordering. A workload mix specifies the probability for each kind of request. In all the experiments reported in this paper, the ordering mix is used because it is the most update-intensive mix (30% of all requests lead to an update). The TPC-W benchmark measures the request throughput by means of emulated browsers (EBs). Each EB simulates one user who issues a request, waits for the answer, and then issues the next request after a specified waiting time. The TPC-W metric for throughput is Web Interactions Per Second (WIPS). According to the TPC-W specification, 90% of requests must meet the response time requirements. Depending on the kind of request, the allowed response time varies from 3 to 20 seconds.

In our experiments, we concentrate on the query execution part of TPC-W. Thus, we do not render the full web pages but only execute the queries to retrieve the data per page. We do not directly compare our query latency values to the given SLOs since they are in terms of end-to-end latency measured at the browser; however, since our query latency values are small compared to the given SLOs (see Table 1), they are clearly not going to be the cause of violations.

Furthermore, while standard TPC-W requires full ACID guarantees, we implement only the semantics described in Section 7.2. Finally, we forego the wait time between requests, allowing us to place more load on the system with fewer machines.

### *8.1.2 SCADr*

SCADr is a website that simulates the microblogging platform Twitter by allowing users to post "thoughts" of at most 140 characters. Users can create a list of other users that they wish to follow, and the most recent thoughts from these users will be displayed in a thoughtstream when they log into the site.

The schema for this application is relatively simple and consists of three tables: users, subscriptions, and thoughts. The users table contains a username as primary key as well as normal user attributes such as password and hometown. The subscriptions table specifies which users are subscribed to whom; that is, it models the n-to-m relationship between the users themselves. The primary key of the subscriptions table is composed of the owner of the subscriptions, followed by the target user. An additional attribute of the table specifies if the subscription has been approved. Finally, the thoughts table stores all the thoughts (i.e., microblog posts) of a user. The thoughts relation is composed of three attributes: username, timestamp, and the actual message, which is limited to 140 characters. The primary key of the thoughts table is composed of the username and the timestamp of the thought.

Our SCADr benchmark defines 5 different kinds of queries: "List users I'm following", "List my recent thoughts", "List the most recent thoughts of all of the people I am subscribed to", "Find user", and finally "Post a new thought", the only updating query. We then measure both the request throughput and response time for executing all queries for a randomly selected user. This workload simulates a group of applications servers issuing database queries against the PIQL system, but not the page rendering portion of the site. Except for the "Post a new thought", which occurs with a probability of 1%, each of the remaining queries is executed once for every simulated request. The "Post a new thought" query is not further considered, as it is just a single put request.

## 8.2 Qualitative Analysis

In this section, we describe how we turn the original SQL queries of the two benchmarks into performance predictable PIQL queries. We optimize each query with the PIQL optimizer and follow the suggestions of the Performance Insight Assistant to rewrite the query where applicable. Table 1 summarizes the necessary modifications (to either the query or the schema) for making the queries scale-independent, as well as the compiler selected indexes.

Although we expected many changes, in particular for the TPC-W queries, surprisingly few changes are required. Most notably, the TPC-W queries require rewriting more general `LIKE` predicates as tokenized keyword searches. This change is an artifact of our current implementation, as we only support inverted full-text indexes for such queries.

The only real change required from the developer is the addition of a cardinality constraint on the number of items inside a shopping cart, though this limit is already defined as an optional constraint in the TPC-W specification. All TPC-W queries except "Best Seller" and "Admin Confirm" are already scale-independent. In addition to the primary keys, the compiler automatically creates 5 indexes to support all queries more efficiently.

We did not implement "Best Sellers" and "Admin Confirm", as both queries are analytical and are best implemented by materialized views even in traditional database systems. Other work [21] also reported ignoring those queries for the scale-up experiment. We plan to address these types of analytical queries using precomputed results as part of our future work. The table also omits the queries whose SQL is identical to that of another (e.g., Search By Subject WI and New Product WI) to save space.

The queries for SCADr require a limit on the number of possible subscriptions per user, similar to how Facebook limits the number of friends, as well as on the number of results shown per page. In our scale experiment, we set the limits to 10 subscriptions and 10 results per page. Refer to Section 6.4 for more detail on how different cardinality limits affect query performance.

## 8.3 Comparison To Cost-Based Optimization

Recall from Section 1 that the PIQL optimizer prioritizes scale-independent plans over cost-optimal plans. As a result, there are cases where PIQL will select a plan that does not execute as quickly as possible over small amounts of data.

In order to quantify this effect, we executed the following query over different amounts of data using both a scale-independent plan and a cost-based one that minimized the average number of operations performed against the key/value store. This query checks to see which of the current user's friends are also subscribed to the user whose profile is being viewed, and is expressed in SQL as:

```
SELECT * FROM SUBSCRIPTIONS
WHERE target = <target user>
AND owner IN <friends of current user>
```

The PIQL compiler selects a plan that performs a bounded number of random read requests against the subscriptions table, checking whether each of the current user's friends exists in the list of the target user's subscribers. In contrast, a cost-based optimizer would decide that since the average number of subscribers for any user is small (in 2009, the average Twitter user had only 126 followers [5]), the best plan is an unbounded index scan which will on average require only one RPC against the key/value store. This plan first retrieves all users who subscribe to the target user, and filters to return only those to whom the current user subscribes.

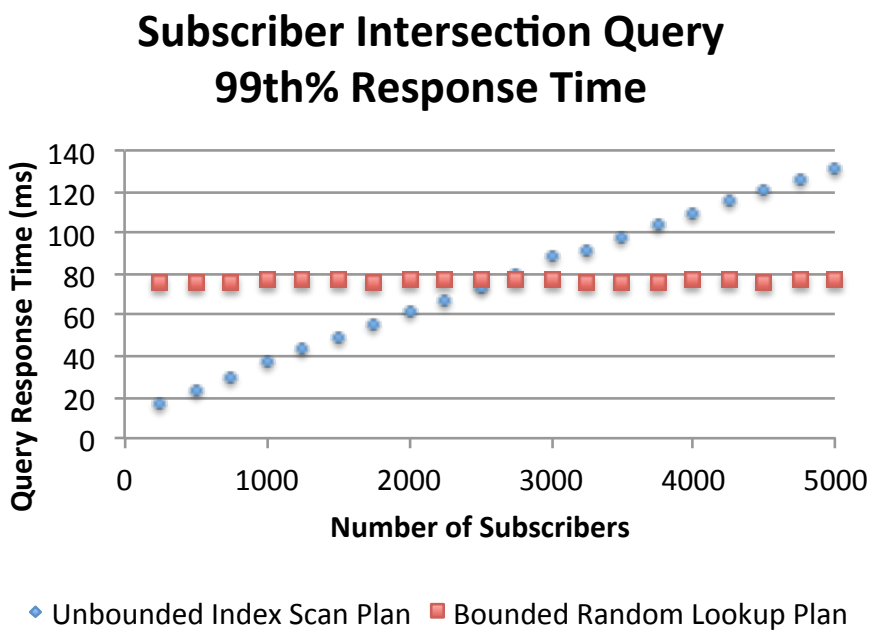

**Figure 7: A comparison of the 99th percentile response time of 200,000 executions of the subscriber intersection query using two different optimization strategies.**

To compare the performance of this query, we ran it against users of increasing popularity, with 50 randomly selected users as the 'friends' of the current user. Figure 7 shows that while the cost-based plan performs up to 4x faster for an unpopular user, the scale-independent plan consistently meets the application's SLO, independent of the popularity of the target user. For a popular user, such as "Lady GaGa" (12M+ followers), using the cost-based query plan would certainly violate the SLO by performing an unbounded index scan to retrieve the people following her. In contrast, PIQL's scale-independent plan would perform a fixed number of random reads, bounded by the cardinality limit on the number of users to whom a given user can subscribe. If it is important to minimize response time, even below the SLO, our system could be extended to use a dynamic approach that determines at runtime which query plan to execute. However, special care would need to be taken to ensure unbounded plans were never run over unsafe amounts of data.

## 8.4 Scale Experiment

In order to evaluate the scale independence, we run both benchmarks on clusters of various sizes and measure the web interaction latency. For each point, we keep the amount of data per server constant while increasing the number of storage nodes and client libraries issuing queries.

### *8.4.1 TPC-W Query Execution*

We scale TPC-W by first bulk loading 75 Emulated Browsers' worth of user data for each storage node in the cluster. The number of items is kept constant at 10,000. Each piece of data is replicated on two servers both for availability and performance reasons.

We then run one client machine with the PIQL library for every two servers in the system, varying the number of storage servers from 20 to 100 (including clients, up to 150 EC2 instances). All data in the system is replicated twice for availability and performance. Each client executes the queries from the workflow specified by the TPC-W benchmark in 10 concurrent threads. Throughput and response time values are collected in 5-minute intervals, with at least 5 iterations for each configuration. We discard the first run of any given setup to avoid performance anomalies caused by JITing and other warm-up effects.

Figures 8 and 9 show the results. The first graph shows a near linear scale-up of throughput as the number of servers and clients increases. At the same time, Figure 9 shows that the response time per web interaction stays virtually constant, even in the 99th percentile, independent of the scale. Thus, PIQL and its execution engine are able to preserve the scalability and predictable performance of the underlying key/value store even for a complicated application like TPC-W. Note, the response times from this experiment are not directly comparable with the predicted response times. This is due to the fact that full web interactions also result in puts to the key/value store, and thus measured response time is slightly higher than predicted total response time.

### *8.4.2 SCADr*

We scale SCADr using a methodology similar to the TPC-W benchmark by varying the number of storage nodes and clients. As with TPC-W, the data size increases linearly with the number of servers, with 60,000 users per server, 100 thoughts per user, and 10 random subscriptions per user. As with TPC-W, all data is replicated on two servers for increased availability.

We then run one client machine with the PIQL library for every two servers in the system, varying the number of storage servers from 20 to 100 (including clients, up to 150 EC2 instances). Each client machine repeatedly simulates the rendering of the "home page" for SCADr by executing all of the given queries and measuring the overall response time. This execution and measurement is done by 10 concurrent threads on each client machine. Throughput and response time statistics are collected in 5-minute intervals, with at least 5 iterations for each configuration. Again, we discard the first run of any given setup to avoid performance anomalies cause by JITing and combine all subsequent response time data to calculate a single 99th percentile value.

Figure 10 and 11 show the results of this experiment. The first graph shows a near linear scale-up of throughput as the number of servers and clients increases. Again, we observe near linear scalability (Figure 10) with low response time, even at the 99th percentile on a public cloud (Figure 11).

## 8.5 Execution Strategies

We also evaluate the effect of different execution strategies on the TPC-W queries' response time. In this experiment, we use three different variants of the PIQL execution engine running on a cluster with 10 storage nodes and 5 client machines. The amount of data and the length of the experiment are kept the same as in the TPC-W scale experiment described in the previous section. Figure 12 shows the 99th-percentile latency for each execution strategy. The first strategy, called the Lazy Executor, operates in a similar fashion to a traditional relational database by requesting a single tuple at a time from the key/value store. The second strategy, called the Simple Executor, utilizes the extra limit hint information provided by optimizer to request data in batches from the key/value store; however, it waits for each request to return before issuing the next. The final strategy, called the Parallel Executor, uses the extra limit hint information and issues all key/value store requests in parallel for a given remote operator. The results show the importance of both the limit hint information and the intra-query parallelism provided by the PIQL execution engine.

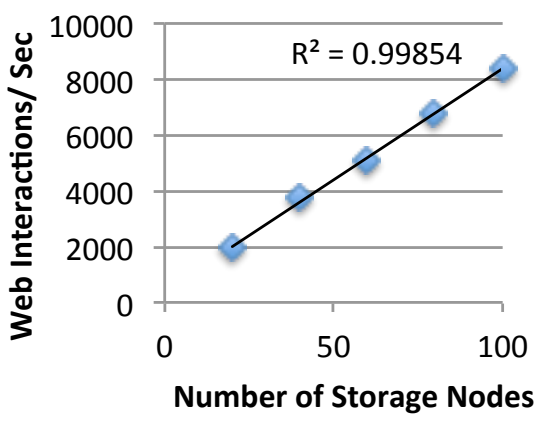

Figure 8: TPC-W Throughput, Varying the Number of Servers

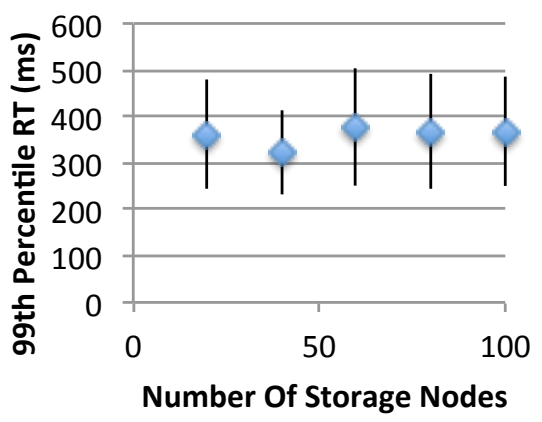

Figure 9: TPC-W 99th Perc. Response Time, Varying the Number of Servers

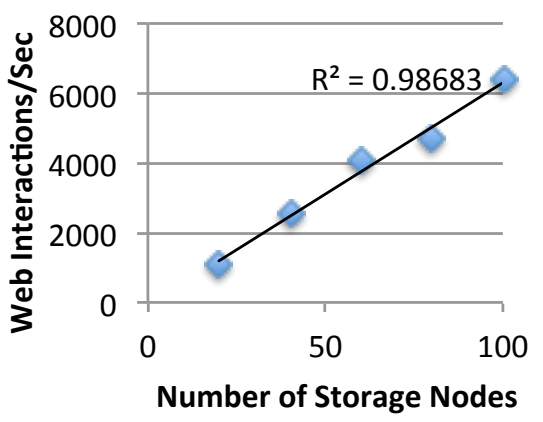

Figure 10: SCADr Throughput, Varying the Number of Servers

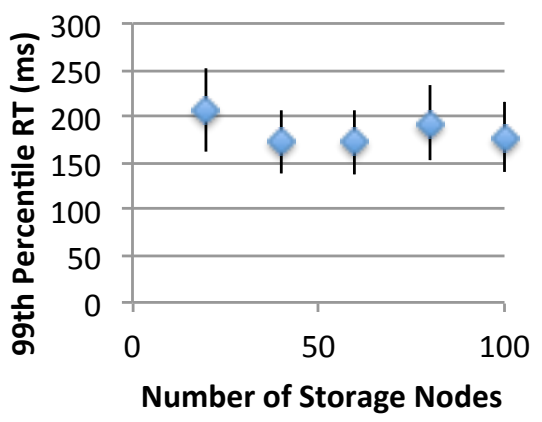

Figure 11: SCADr 99th Perc. Response Time, Varying the Number of Servers

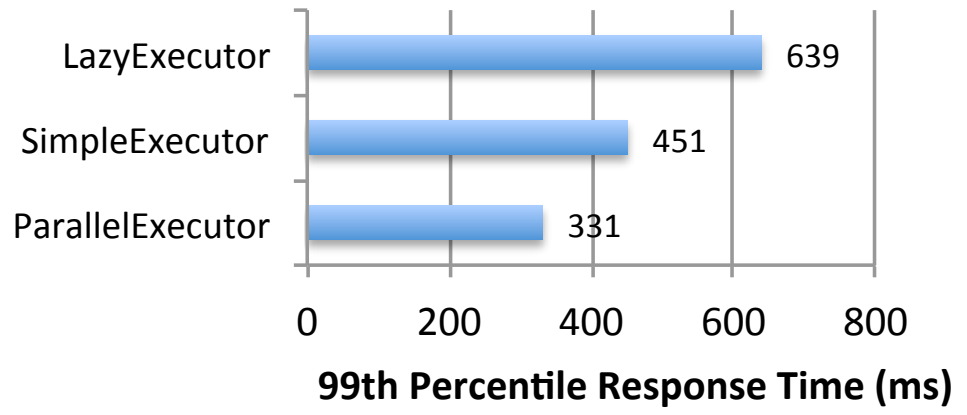

Figure 12: TPC-W 99th Perc. Response Time, Varying the Execution Strategy

### 8.6 Prediction

Table 1 shows the actual and predicted 99th-percentile values for each of the queries in TPC-W and SCADr. For brevity, we maintain a constant cardinality for each query; for predictions with a varying cardinality, see Section 6.4.

To train our model, we benchmark the operators on Amazon EC2 with a 10-node cluster using two-fold replication and retrieve statistics for 35 10-minute intervals. We obtain a prediction of the per-query 99th-percentile latency for each interval as described in Section 6.3. In the table, for both the actual and predicted cases, we report the max 99th-percentile value. Taking the max corresponds to a very conservative approach to setting the cardinality.

As the results show, we slightly overestimate the actual 99th-percentile value in most cases. As we mentioned in Section 6, our goal is to predict not response time but rather SLO compliance; thus, we prefer to over-predict as long as the difference between the predicted and actual values is not so large as to be untrustworthy. We do underestimate three queries by 2 ms, but we consider this to be insignificant for the purposes of determining SLO compliance.

Our model is most conservative for TPC-W's New Products WI. This overestimation is caused by pipelining between the query's two index foreign key joins, which our model does not currently capture. In the future, we plan to extend our model to handle this case. If the SLO is sufficiently above our predicted value, the developer will still be able to make the correct decision regarding SLO compliance. However, a consequence of this modeling error is that the Performance Insight Assistant described in Section 6.4 could potentially recommend a cardinality value lower than what the system could actually handle while still meeting its SLO. Thus, developers should take the assistant-recommended cardinality as a starting point and potentially increase the cardinality over time if the performance of the deployed application seems to consistently be well below the SLO.

The SLOs provided with the TPC-W specification, which range from 3 to 5 seconds for the queries we considered, are in terms of the 90th-percentile end-to-end response time, while we look at the 99th-percentile query response time. Therefore, they are not directly comparable with our results. However, the running time of our queries, even at the 99th percentile, is much less than the given SLOs; clearly, the queries would not be the bottleneck to meeting the SLOs. We do not have any SLOs provided for SCADr, since we devised the benchmark. We chose 500 ms as our target latency since longer server delays have been shown to affect the number of queries performed by a user [26]. Our queries all complete within this bound even for the worst-case 99th-percentile response time.

## 9. RELATED WORK

PIQL combines several database techniques to provide "success-tolerant" query processing. The overall design of PIQL is inspired by the success of key/value stores, such as Dynamo [13] or BigTable [10], and their open-source counterparts like Cassandra [24] and HBase [1]. The architecture PIQL employs is similar to the one proposed in [8], where a key/value store, Amazon S3, is used as a shared disk. However, that work concentrates on consistency and transaction management and does not cover the query language and query execution. PIQL leverages get and put operations with predictable performance gained from techniques described by Trushkowsky et al. [28] to provide a high-level declarative query language which preserves this predictability.

The query language and the optimization techniques of PIQL are closely related to top-k query processing. Carey and Kossmann presented the concept of stop-after operators and demonstrated how they could be used to efficiently execute relational top-k queries [9]. PIQL extends the notion of stop operators to include the data-stop operator (Section 5.1) for every constraint on relationship cardinality. This knowledge allows the optimizer to push down the data-stop operator closer to the index scans. As a result, this information enables the calculation of a bound on the number of key/value store operations to be executed for queries where this computation would otherwise be impossible.

The work of [9] has been extended in several directions [20], but the objectives of these techniques differ significantly from the design goals of PIQL. Most prominently, all top-k techniques strive to minimize query-processing time, whereas we effectively want to bound the response time. Still, some of the ideas, like threshold algorithms (TA) from Fagin et al. [14], or pipelined rank-join operator from Ilyas et al. [19], could help to decrease response times.

Ganapathi et al. [15] achieve good prediction for several query metrics, including latency and resource consumption, using a black-box technique. In contrast to our approach, this technique requires the training set to contain queries that are similar to the new query. Furthermore, this technique has only been validated when each query is run in isolation on the database.

In addition to cost models for query optimization, many latency models for web-service requests have been preposed [29]. Recent work has successfully modeled high-latency quantiles of RUBiS2 transactions in order to assess the impact of CPU allocation on response time in a virtualized environment [30].

The WSQ/DSQ system [16] reduces overall query latency when retrieving information from a high latency store, in their case a search engine, through a technique called asynchronous iteration. Similarly, the PIQL execution engine leverages parallelism to execute queries more quickly against the underlying key/value store.

# 10. CONCLUSION

Guaranteeing good performance for interactive web applications is more important than ever. While it is easy to develop fast web sites when the user base is small, it is hard to guarantee that the performance will be acceptable at scale.

With PIQL, we introduced a new form of data independence, called scale independence, to ensure that queries that perform well on a small amount of data will continue to meet the SLO requirements as the database size grows. We described the PIQL language along with its accompanying compiler and SLO prediction framework, which allow analyzing queries for their scale independence and SLO compliance at compile time. We further showed how PIQL's architecture and execution engine leverages a key/value store as a record manager to achieve high scalability with predictable performance. Finally, we demonstrated the feasibility of our approach by an extensive evaluation of the expressiveness of our language, the accuracy of our prediction model, as well as the performance and scaling properties of the PIQL system using the TPC-W and a Twitter-like benchmark.

There are several interesting future directions for PIQL. While PIQL currently uses a simple rule-based optimizer, the prediction model is already the first step towards a cost-based optimizer. Hence, it seems natural to explore the model for cost-based plan selections. Also, some useful queries are currently disallowed by PIQL due to the constraint that the number of operations needs to have a compile time upper-bound. One possible way to relax this constraint is to relax the freshness requirements on the query and instead answer queries through pre-computation. Finally, it would be interesting to develop a formal model of scale independence which could be used to determine which SQL features are inherently unscalable.

By adding scale independence to the relational model, PIQL combines the performance predictability and scalability of key/value stores with the convenience of a high-level declarative language, enabling many more web applications to achieve success tolerance.

# 11. ACKNOWLEDGMENTS

We would like to thank the following people, along with the reviewers, for feedback that significantly improved the content of this paper: Bill Bolosky, Henry Cook, Michael Jordan, Kimberly Keeton, Nick Lanham, Beth Trushkowsky, and Stephen Tu. This research is supported in part by gifts from Google, SAP, Amazon Web Services, Cloudera, Ericsson, Huawei, IBM, Intel, MarkLogic, Microsoft, NEC Labs, NetApp, Oracle, Splunk, and VMware and by DARPA (contract #FA8650-11-C-7136).